\documentclass[journal,10pt]{IEEEtran}
\usepackage{mathtools}
\usepackage{amsmath}
\usepackage{algorithmic}
\usepackage{algorithm}
\usepackage{graphicx, graphics, theorem, times, amsfonts, amsmath, amssymb, cite}
\usepackage{pgfplots}
\usepackage{color}
\usepackage{multirow}
\usepackage{rotating}
\usepackage{bm}

\newcounter{excercise}
\newcounter{excercisepart}

\definecolor{pennblue}{cmyk}{1,0.65,0,0.30}
\definecolor{pennred}{cmyk}{0,1,0.65,0.34}
\definecolor{mygreen}{rgb}{0.10,0.50,0.10}

\def \diag    {\text{\normalfont diag} }

\def \reals    {{\mathbb R}}

\newcommand{\argmin}{\operatornamewithlimits{argmin}}

\def\ccalG{{\ensuremath{\mathcal G}}}

\def\ccalI{{\ensuremath{\mathcal I}}}

\def\ccalV{{\ensuremath{\mathcal V}}}
\def\ccalS{{\ensuremath{\mathcal S}}}

\def\ccal0{{\ensuremath{\mathcal 0}}}
\def\bbA{{\ensuremath{\mathbf A}}}

\def\bbD{{\ensuremath{\mathbf D}}}

\def\bbG{{\ensuremath{\mathbf G}}}
\def\bbH{{\ensuremath{\mathbf H}}}
\def\bbI{{\ensuremath{\mathbf I}}}

\def\bbP{{\ensuremath{\mathbf P}}}

\def\bbR{{\ensuremath{\mathbf R}}}

\def\bbV{{\ensuremath{\mathbf V}}}
\def\bbS{{\ensuremath{\mathbf S}}}

\def\bbX{{\ensuremath{\mathbf X}}}
\def\bbY{{\ensuremath{\mathbf Y}}}
\def\bbZ{{\ensuremath{\mathbf Z}}}
\def\bbb{{\ensuremath{\mathbf b}}}

\def\bbe{{\ensuremath{\mathbf e}}}
\def\bbf{{\ensuremath{\mathbf f}}}
\def\bbg{{\ensuremath{\mathbf g}}}
\def\bbh{{\ensuremath{\mathbf h}}}

\def\bbp{{\ensuremath{\mathbf p}}}

\def\bbw{{\ensuremath{\mathbf w}}}
\def\bbu{{\ensuremath{\mathbf u}}}
\def\bbv{{\ensuremath{\mathbf v}}}

\def\bbx{{\ensuremath{\mathbf x}}}
\def\bby{{\ensuremath{\mathbf y}}}

\def\bb0{{\ensuremath{\mathbf 0}}}
\def\hbX{{\hat{\ensuremath{\mathbf X}} }}

\def\tbg{{\tilde{\ensuremath{\mathbf g}} }}
\def\tbh{{\tilde{\ensuremath{\mathbf h}} }}

\def\tbx{{\tilde{\ensuremath{\mathbf x}} }}
\def\tby{{\tilde{\ensuremath{\mathbf y}} }}

\def\bbLambda{\boldsymbol{\Lambda}}

\def\bbPsi{\boldsymbol{\Psi}}

\graphicspath{ {figure/} }

\newcommand{\QED}{\hfill\ensuremath{\blacksquare}}

\newtheorem{myproposition}{\hspace{-1pt}\bf Proposition}
\newtheorem{remark}{\hspace{-1pt}\bf Remark}

\newenvironment{myproof2}[1][$\!\!$]{{\noindent\bf Proof #1: }}
{{\hfill$\blacksquare$\medskip}}

\title{Blind Identification of Invertible Graph Filters \\ with Multiple Sparse Inputs}

\author{
Chang Ye, Rasoul Shafipour and Gonzalo Mateos \\
Dept. of Electrical and Computer Engineering, University of Rochester, Rochester, NY, USA
\thanks{Work in this paper was supported by the NSF award CCF-1750428. Author emails: \{cye7,rshafipo,gmateosb\}@ur.rochester.edu.}}

\begin{document}
%\ninept
%
\maketitle
\thispagestyle{empty}
\begin{abstract}
This paper deals with problem of blind identification of a graph filter and its sparse input signal, thus broadening
the scope of classical blind deconvolution of temporal and spatial
signals to irregular graph domains. While the observations are bilinear functions of the unknowns, a mild requirement on invertibility of the filter enables an efficient convex formulation, without relying on matrix lifting that can hinder applicability to large graphs.  On top of scaling, it is argued that (non-cyclic) permutation ambiguities may arise with some particular graphs.  Deterministic sufficient conditions under which the proposed convex relaxation can exactly recover the unknowns are stated, along with those guaranteeing identifiability under the Bernoulli-Gaussian model for the inputs. Numerical
tests with synthetic and real-world networks illustrate the merits
of the proposed algorithm, as well as the benefits of leveraging multiple
signals to aid the (blind) localization of sources of diffusion.
\end{abstract}
%%%%%%%%%%%%%%%%%%%%%%%%%%%%%%%%%%%%%%%%%%%
%%%%%%%%%%%%%%%%%%%%%%%%%%%%%%%%%%%%%%%%%%%
\section{{Introduction}}\label{S:introduction}
Network processes such as neural activities at different regions of the brain \cite{weiyu_brain_signals,hu2016localizing}, vehicle trajectories over road networks \cite{deri2016new}, or spatial temperature
profiles measured by a wireless sensor network~\cite{D_LMS_TSP}, can be represented as signals supported on the nodes of a graph. Under the natural assumption that the signal properties are influenced by the graph topology (e.g., in a network diffusion or percolation process), the goal of graph signal processing (GSP) is to develop algorithms that exploit this relational structure. Accordingly, generalizations of fundamental signal processing tasks have been widely explored in recent work; see~\cite{gsp2018tutorial} for a comprehensive tutorial treatment. Notably graph filters -- which generalize classical time-invariant systems -- were conceived as information-processing operators acting on graph-valued signals~\cite{sandryhaila2013discrete}. Mathematically, graph filters are linear transformations that can be expressed as
polynomials of the so-termed graph-shift operator (Section \ref{S:prelim}). %\blue{The graph shift gives an alegbraic representation of network structure and can be viewed as a local diffusion operator. For the directed cycle graph representing periodic temporal signals for example, the graph shift boils down to the classical time-shift operator~\cite{sandryhaila2013discrete}.} 
The graph shift offers an alegbraic representation of network structure and can be viewed as a local diffusion operator. For the directed cycle graph representing e.g., periodic temporal signals, it boils down to the classical time-shift operator~\cite{sandryhaila2013discrete}.
Given a shift, the polynomial coefficients fully determine the graph filter and are referred to as filter coefficients. %When graph filters are used to model diffusion processes, the coefficients can be interpreted as diffusion rates. 

\noindent \textbf{Problem outline and envisioned applications.} In this paper, we revisit the blind identification of graph filters with sparse inputs, with emphasis on modeling diffusion processes and localization of the sources of diffusion~\cite{segarra2017blind}. Specifically, given observations of graph signals $\{\bby_i \}_{i=1}^{P}$ that we model as outputs of a diffusion filter (i.e., a polynomial in a known graph-shift operator), we seek to jointly identify the filter coefficients $\bbh$ and the input signals $\{\bbx_i \}_{i=1}^P$ that gave rise to the network observations. This inverse problem broadens the scope of classical blind deconvolution of temporal or spatial signals to graphs \cite{ahmed2014blind,levin2011understanding}. Since the resulting bilinear inverse problem is ill-posed, we assume that the inputs are sparse -- a well-motivated setting when few seeding nodes inject a signal that is diffused throughout a network~\cite{segarra2017blind}. Accordingly, envisioned application domains include environmental monitoring (where are the heat or seismic sources?), opinion formation in social networks (who started the rumor?), neural signal processing (which brain regions were activated?), and epidemiology (who is patient zero for the disease outbreak?). 

\noindent \textbf{Related work and contributions.} Different from most existing works dealing with source localization on graphs, e.g.,  \cite{zhang2016towards,pinto2012locating}, like~\cite{pena2016source} the advocated GSP approach is applicable even when a single snapshot of the diffused signal is available. Often the models of diffusion are probabilistic in nature, and resulting maximum-likelihood source estimators can only be optimal for particular (e.g., tree) graphs~\cite{pinto2012locating}, or rendered
scalable under restrictive dependency assumptions~\cite{feizi2016network}. Relative to~\cite{pena2016source,hu2016localizing}, the proposed framework can accommodate signals defined on general undirected graphs and relies on a convex estimator of the sparse sources of diffusion. Furthermore, the setup where multiple output signals are observed (each one corresponding to a
different sparse input), has not been thoroughly explored in convex-relaxation approaches to blind deconvolution
of (non-graph) signals, e.g.,~\cite{ahmed2014blind,LingBiConvexCS}; see \cite{wang2016blind} for a recent and inspiring alternative that we leverage here. 

A noteworthy approach was put forth in~\cite{segarra2017blind}, which casts the (bilinear) blind graph-filter identification task as a linear inverse problem in the ``lifted'' rank-one, row-sparse matrix $\bbx \bbh^{T}$. While the rank and sparsity minimization algorithms in \cite{segarra2017blind,david_blind} can successfully recover sparse inputs along with low-order graph filters, reliance on matrix lifting can hinder applicability to large graphs. Beyond this computational consideration, the overarching assumption of~\cite{segarra2017blind} is that the inputs $\{\bbx_i \}_{i=1}^P$ share a common support. Here instead we show how a mild requirement on invertibility of the graph filter facilitates an efficient convex formulation for the multi-signal case with arbitrary supports (Section \ref{S:blind_ID}); see also \cite{wang2016blind} for a time-domain precursor. In Section~\ref{S:amb_uni} we take a closer look at inherent scaling and (non-cyclic) permutation ambiguities arising with some particular graphs. We also briefly comment on identifiability under the Bernoulli-Gaussian model for the inputs~\cite{li2015unified}, and state deterministic sufficient conditions under which the proposed convex relaxation can exactly recover the unknowns.
Numerical tests with synthetic graphs and a structural brain network corroborate the effectiveness of the proposed approach in recovering the sparse input signals (Section~\ref{S:simulation}).  Concluding remarks are given in Section~\ref{S:conclusion}.

\section{Preliminaries and Problem Statement} \label{S:prelim}
Consider a weighted and undirected network graph $\ccalG=(\ccalV,\bbA)$, where $\ccalV$ is the set of vertices with cardinality $\lvert \ccalV \rvert=N$, and $\bbA \in \reals^{N \times N}$ is the symmetric graph adjacency matrix whose entry $A_{ij}$ denotes the edge weight between nodes $i$ and $j$. As a more general algebraic descriptor of network structure, one can define a \textit{graph-shift operator} $\bbS \in \reals^{N \times N}$ as any matrix having the same sparsity pattern as $\ccalG$ \cite{sandryhaila2013discrete}. Accordingly, $\bbS$ can be viewed as a local diffusion (or averaging) operator. Common choices are to set it to either $\bbA$ (and its normalized counterparts) or variations of adjacency and Laplacian matrices \cite{gavili2017shift,gsp2018tutorial}. Since $\bbS$ is real and symmetric, it is diagonalizable so that $\bbS=\bbV\bbLambda\bbV^T$, with $\bbLambda=\textrm{diag}(\lambda_1,\ldots,\lambda_N)$. Lastly, a graph signal $\bbx: \ccalV  \mapsto \reals^N$ is an $N$-dimensional vector, where entry $x_i$ represents the signal value at node $i \in \ccalV$. 

\vspace*{-0.3cm}
\subsection{Graph-filter models of network diffusion processes}
%\noindent\textbf{Graph-filter models of network diffusion processes.} 
Let $\bby$ be a graph signal supported on $\ccalG$, which is generated from an input graph signal $\bbx$ via linear network dynamics of the form
\begin{align}\label{eqn_diffusion}
	\textstyle \bby\  =\ \alpha_0 \prod_{l=1}^{\infty} (\bbI-\alpha_l \bbS) \bbx
	\  =\ \sum _{l=0}^{\infty} \beta_l \bbS^l \bbx .
	%   \ :=\  \bbH \bbw.
\end{align}
While $\bbS$ encodes only one-hop interactions, each successive application of the shift in \eqref{eqn_diffusion} diffuses $\bbx$ over $\ccalG$. %The product and sum representations in \eqref{eqn_diffusion} are common (and equivalent) models for the generation of linear network processes. 
Indeed, any process that can be understood as the linear propagation of a seed signal through a static graph can be written in the form in \eqref{eqn_diffusion}, and subsumes heat diffusion, consensus and the classic DeGroot model of opinion dynamics as special cases~\cite{DeGrootConsensus}.

The diffusion expressions in \eqref{eqn_diffusion} are polynomials on $\bbS$ of possibly infinite degree, yet the Cayley-Hamilton theorem asserts they are equivalent to polynomials of degree smaller than $N$. Upon defining the vector of coefficients $\bbh:=[h_0,\ldots,h_{L-1}]^T$ and the shift-invariant graph filter
\begin{equation}\label{e:graph_filter_def}
	\mathbf{H}:=h_0\bbI_N+h_1 \mathbf{S}+h_2 \mathbf{S}^2+\ldots+h_{L-1} \mathbf{S}^{L-1}=\sum_{l=0}^{L-1}h_l \mathbf{S}^l,
\end{equation}
the signal model in \eqref{eqn_diffusion} becomes
$\bby  = \big(\sum_{l=0}^{L-1}h_l \bbS^l\big)\,\bbx
:= \bbH \bbx$, for some particular $\bbh$ and $L\leq N$. Due to the local structure of $\bbS$, graph filters represent linear transformations that can be implemented in a distributed fashion~\cite{segarra2015distributed}, e.g., via
$L-1$ successive exchanges of information among neighbors. %Since $\bbH$ is a polynomial on $\bbS$~\cite{sandryhaila2013}, they are linear graph-signal operators that have the same eigenvectors as the shift (i.e., the operators $\bbH$ and $\bbS$ commute and hence they represent \emph{shift-invariant} transformations). 

Leveraging the spectral decomposition of $\bbS$, graph filters and signals can be represented in the frequency domain.
Specifically, let us use the eigenvalues of $\bbS$ to define the $N\times L$ Vandermonde matrix  $\bbPsi_L$, where $\Psi_{ij}:=\lambda_i^{j-1}$.
The frequency representations of a signal $\bbx$ and filter $\bbh$ are defined as $\tbx:=\bbV^T\bbx$ and $\tbh:=\bbPsi_L\bbh$, respectively. The latter  follows since the  output $\bby\!=\!\bbH\bbx$ in the frequency domain is given by
\begin{equation} \label{e:freq_response}
\tby=\diag\big(\bbPsi_L\bbh\big)\bbV^T \bbx=\diag\big(\tbh\big)\tbx=\tbh\circ \tbx.
\end{equation} 
This identity can be
seen as a counterpart of the convolution theorem for temporal signals, where $\tby$ is the elementwise product $(\circ)$ of $\tbx$ and the filter's frequency response $\tbh:=\bbPsi_L\bbh$.

\subsection{Problem formulation}
%\noindent\textbf{Problem formulation.} 
For given shift operator $\bbS$ and filter order $L$, suppose we observe $P$ output signals collected in a matrix $\bbY=[\bby_1,\ldots,\bby_P] \in \reals^{N \times P}$ such that $\bbY = \bbH \bbX$, where $\bbX = [\bbx_1,\ldots,\bbx_P]\in \reals^{N \times P}$ is sparse having at most $S\ll N$ non-zero entries per column. The goal is to perform blind identification of the graph filter (and its input
signals), which amounts to estimating sparse $\bbX$ and the filter coefficients $\bbh$ up to scaling and (possibly) permutation ambiguities; see Section \ref{S:amb_uni}. Sparsity is well motivated when the signals in $\bbY$ represent diffused versions of a few localized sources in $\ccalG$, here indexed by $\textrm{supp}(\bbX):=\{(i,j) \mid X_{ij} \neq 0 \}$. Moreover, the non-sparse formulation is ill-posed, since the number of unknowns $NP + L$ in $\{\bbX,\bbh\}$ exceeds the $NP$  observations  in $\bbY$. %Alternatively, a low-dimensional subspace model for $\bbX$ can be adopted to effectively reduce the degrees of freedom in the problem \cite{segarra2017blind}. 

All in all, using \eqref{e:freq_response} the diffused source localization task can be stated as a feasibility problem of the form
\begin{equation} \label{e:blind_feasibility}
\text{find } \{ \bbX,\bbh \} \:\: \text{s. to }\: \bbY = \bbV\diag\big(\bbPsi_L\bbh\big)\bbV^T\bbX, \: \| \bbX \|_{0} \leq PS,
\end{equation}
where the $\ell_0$-(pseudo) norm $\| \bbX \|_{0}:=|\textrm{supp}(\bbX)|$ counts the non-zero entries in $\bbX$. In words, the goal is to find the solution to a system of bilinear equations subject to a sparsity constraint in $\bbX$; a hard problem due to the non-convex $\ell_0$-norm as well as the bilinear constraints. To deal with the latter, building on~\cite{wang2016blind} we will henceforth assume that the filter $\bbH$ is invertible. %But before moving on, a review of related GSP concepts is in order.

\section{Convex Relaxation for Invertible Filters}\label{S:blind_ID}
Here we show how to efficiently tackle the blind graph filter identification problem, through a convex relaxation of \eqref{e:blind_feasibility} when the diffusion filter is invertible.  

To that end, note from \eqref{e:freq_response} that graph filter $\bbH$ is invertible if and only if $\tilde{h}_i=\sum_{l=0}^{L-1} h_l \lambda_{i}^l \neq 0$, for all  $i=1,\ldots,N$. In words, the frequency response of the filter should not vanish at the graph frequencies $\{\lambda_i\}$. In such case one can show that the inverse operator $\bbG := \bbH^{-1}$ is also a graph filter on $\ccalG$, which can be uniquely represented as a polynomial in the shift $\bbS$ of degree at most $N-1$ \cite[Theorem 4]{sandryhaila2013discrete}. To be more specific, let $\bbg \in \reals^{N}$ be the vector of inverse-filter coefficients, i.e., $\bbG= \sum_{l=0}^{N-1} g_l \bbS^l$. Then one can equivalently rewrite the generative model $\bbY=\bbH\bbX$ for the observations as
\begin{equation} \label{e:Filter_G}
		\bbX  = \bbG \bbY  = \bbV \text{diag}(\tbg) \bbV^T \bbY,
\end{equation}
where $\tilde{\bbg} := \bbPsi_N \bbg \in \reals^N$ is the inverse filter's frequency response and $\bbPsi_N \in \reals^{N \times N}$ is Vandermonde. Naturally, $\bbG=\bbH^{-1}$ implies the condition $\tbg\circ\tbh =\mathbf{1}_N$ on the frequency responses, where $\mathbf{1}_N$ denotes the $N \times 1$ vector of all ones. Leveraging \eqref{e:Filter_G}, one can recast \eqref{e:blind_feasibility} as a \emph{linear} inverse problem
\begin{equation} \label{e:opt_blind_invertible}
 \min_{\{\bbX,\tbg \}}
\:\| \bbX \|_{0},\:\: \text{s. to }\:
\bbX = \bbV \text{diag}(\tbg) \bbV^T \bbY,\:\bbX\neq\mathbf{0}.
\end{equation}
This approach is markedly different from the matrix lifting technique used in~\cite{segarra2017blind} to handle the bilinear equations in \eqref{e:blind_feasibility}.

The $\ell_0$ norm in \eqref{e:opt_blind_invertible} makes the problem NP-hard to optimize. Over the last decade or so, convex-relaxation
approaches to tackle sparsity minimization problems have enjoyed
remarkable success, since they often
 entail no loss of optimality. Accordingly, we instead: (i) seek to minimize the $\ell_1$-norm convex surrogate of the cardinality function, that is $\| \bbX \|_{1} = \sum_{i,j}| X_{ij}|$; and (ii) express the filter in the graph spectral domain as in \eqref{e:Filter_G} to obtain the cost 
%
%\begin{align*}
%\| \bbX \|_{1} ={}& \| \bbG \bbY \|_{1}\\
%={} & \| \bbV \text{diag}(\tilde{\bbg}) \bbV^T \bbY \|_{1}\\
%={}&\| (\bbY^{T}\bbV \odot \bbV) \tilde{\bbg} \|_{1},
%\end{align*}
\begin{equation*}
	\| \bbX \|_{1} = \| \bbG \bbY \|_{1} 
	= \| \bbV \text{diag}(\tilde{\bbg}) \bbV^T \bbY \|_{1}
	=\| (\bbY^{T}\bbV \odot \bbV) \tilde{\bbg} \|_{1},
\end{equation*}
where $\odot$ denotes the Khatri-Rao (i.e., columnwise Kronecker) product. This suggests solving the convex $\ell_1$-synthesis problem (in this case a linear program), e.g.,~\cite{zhang2016one}, namely
\begin{equation} \label{e:opt_prob_convex}
\widehat{\tilde{\bbg}}= \argmin_{\tilde{\bbg} \in \reals^{N}}
\:\| (\bbY^{T}\bbV\odot \bbV) \tilde{\bbg} \|_{1},\quad \text{s. to }\:
\mathbf{1}_N^{T} \tilde{\bbg} = 1.
\end{equation}
While the linear constraint in \eqref{e:opt_prob_convex} avoids $\widehat{\tilde{\bbg}} = 0$, it also serves to fix the scale of the solution.

As a result, under the pragmatic assumption that the diffusion filter is invertible, one can readily use e.g., an off-the-shelf interior-point method or a specialized sparsity-minimization algorithm to solve \eqref{e:opt_prob_convex} efficiently. Different from the solvers in~\cite{segarra2017blind,david_blind}, the aforementioned algorithmic alternatives are free of expensive singular-value decompositions per iteration. 
We have found that overall performance can be improved via the iteratively-reweighted $\ell_1$-norm minimization procedure tabulated under Algorithm \ref{alg:reweighted_cvx}; see also~\cite{candes2008enhancing} for a justification of such refinement. In any case,  notice that once the frequency response $\widehat{\tbg}$ of the inverse filter is recovered, one can readily reconstruct the sources via $\hbX =  (\bbY^{T}\bbV \odot \bbV) \tbg$ as well as the filter $\bbH$, if desired.

\begin{algorithm}[t]
	\caption{Iteratively-reweighted $\ell_1$ minimization for \eqref{e:opt_prob_convex}}
	\label{alg:reweighted_cvx}
	\begin{algorithmic}[1]
		\STATE 	\textbf{Input: } Matrix $\bbY^{T}\bbV \odot \bbV$, $\delta > 0$ and  $\epsilon > 0$.
		\STATE \textbf{Initialize} $t=0$, $\bbw^{(0)} = \mathbf{1}_{NP}$ and $\bbX^{(0)} = \mathbf{0}$.
		\REPEAT
		\STATE Solve \begin{equation} \begin{aligned} \nonumber \tilde{\bbg}^{(t+1)}  = & \argmin_{\tilde{\bbg}} \left\| \bbw^{(t)}\circ \big [ (\bbY^{T}\bbV \odot \bbV) \tilde{\bbg}\big ]\right\|_{1} \\  &\text{s. to } \quad \mathbf{1}_{N}^T\tilde{\bbg} = 1. \end{aligned} \end{equation}
		\STATE Form $\bbX^{(t+1)} = (\bbY^{T}\bbV \odot \bbV)\tilde{\bbg}^{(t+1)}$.
		\STATE Update $w^{(t+1)}_i = \frac{1}{[\text{vec}(\bbX^{(t+1)})]_i + \delta}, \quad i = 1,2,...,NP$.
		\STATE $t \gets t+1$.
		\UNTIL $\Vert\bbX^{(t+1)} - \bbX^{(t)}\rVert_{1} / \Vert\bbX^{(t)}\rVert_{1} \le \epsilon$
		\RETURN $\hat{\tilde{\bbg}}:= \tilde{\bbg}^{(t+1)}$ and $\hat{\bbX}:=\bbX^{(t+1)}$
	\end{algorithmic}
\end{algorithm}

In the next section we will take a closer look at the inherent  ambiguities associated with the bilinear model $\bbY=\bbH\bbX$, some of which are unique to the network setting dealt with here. These are of course important to delineate the scope of identifiability (i.e., uniqueness) results.  We will complete our discussion with deterministic sufficient conditions under which the convex relaxation~\eqref{e:opt_prob_convex} is tight.

\section{Identifiability and Exact Recovery}\label{S:amb_uni}
To establish further connections with blind deconvolution of periodic discrete-time signals,  recall these can be
viewed as graph signals supported on the directed cycle graph (whose circulant adjacency matrix is diagonalized by the DFT basis)~\cite{sandryhaila2013discrete}. In this special case, the blind identification task is known to suffer from unavoidable scaling and circulant-shift ambiguities; see e.g.,~\cite{ahmed2014blind,wang2016blind}. Here we examine more general symmetric permutation ambiguities arising with unweighted graphs, and briefly outline a relevant identifiability result as well as preliminary exact recovery conditions for \eqref{e:opt_prob_convex}.

\begin{figure}[t]
	\centering    
	{\includegraphics[width=1\linewidth]{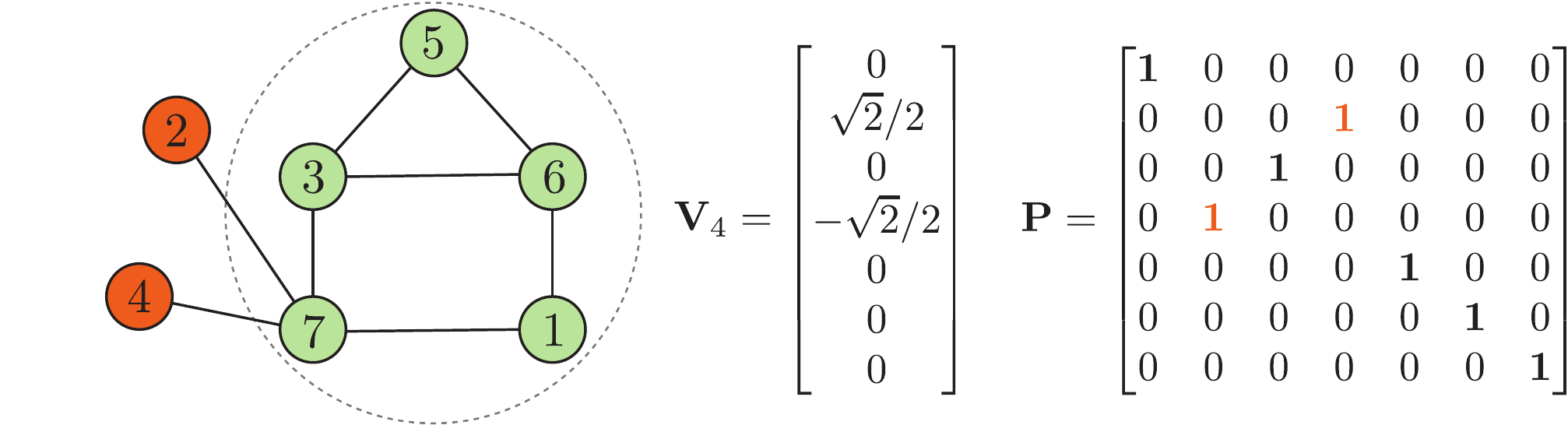}}
	\caption{Toy undirected graph (left) used to illustrate the symmetric permutation ambiguity between nodes $2$ and $4$. The fourth eigenvector $\bbv_4$ of $\bbS=\bbA$ (center) has the problematic form. Then if $\{\bbX_0,\tilde{\bbh}_0 \}$ satisfies the bilinear equations $\bbY = \bbV\textrm{diag}(\tbh)\bbV^T \bbX$, so does $\{ \bbP \bbX_0
		, \text{diag}(\bbp) \tilde{\bbh}_0\}$ for the shown permutation matrix $\bbP$ (right) and $\bbp = [1,1,1,-1,1,1,1]^{T}$.}
	\label{fig:example_ambiguity}
	\vspace{-0.3cm}
\end{figure}

\vspace*{-0.3cm}
\subsection{Permutation ambiguities for some unweighted graphs} \label{S_s:per}
In solving the bilinear inverse problem formulated in Section \ref{S:prelim}, for some particular graphs in addition to scaling we may also encounter (non-cyclic shift) permutation ambiguities. We can resolve the scaling ambiguity by e.g., a fortiori setting $\|  \tilde{\bbg}_{0}\|_{1}=1$ as in the experiments of Section~\ref{S:simulation}, where $\tilde{\bbg}_{0}$ is the ground-truth frequency response of the inverse filter. %However, identifying the permutation ambiguity is in general challenging. 
Inspired by the identifiability studies for sparsity-constrained bilinear problems~\cite{li2015unified}, here we examine said permutation ambiguities for unweighted graphs with shift $\bbS = \bbV \bbLambda \bbV^T$. 

Let  $\{\bbX_0,\tilde{\bbh}_0 \}$ collect the ground-truth sparse input signals and the filter's frequency response, respectively. Let $\bbu^{(i,j)} \in \mathbb{R}^N$ be a unit-norm vector with zero entries except for $u^{(i,j)}_{i} = -u^{(i,j)}_{j} = 1/\sqrt{2}$. As we show next, a permutation ambiguity arises if, say, the $k$th eigenvector of $\bbS$ (i.e., the $k$th column of $\bbV$) has the form $\bbu^{(i,j)}$. Indeed, in that case one could introduce a binary signed vector $\bbp \in \{-1,1\}^N$ with a single negative entry $p_k=-1$, to construct another solution of the form
\begin{equation} \label{e:alternative_solution}
\bbX_1 := \bbP \bbX_0,\quad\tilde{\bbh}_1  := \text{diag}(\bbp) \tilde{\bbh}_0,
\end{equation}
where
%
%\begin{equation} \label{e:permutation_decomposition}
$\bbP  = \mathbf{I}_N - 2 \bbu^{(i,j)}(\bbu^{(i,j)})^T = \bbV\text{diag}(\bbp)\bbV^T$
%\end{equation}
%
%Operator $\bbP$ in \eqref{e:permutation_decomposition} 
is a symmetric permutation matrix that interchanges the signal values at nodes $i$ ($x_i$) and $j$ ($x_j$) when applied to the graph signal $\bbx$. It is immediate that the pair in \eqref{e:alternative_solution} satisfies the generative model $\bbY = \bbH \bbX = \bbV \text{diag}(\tbh) \bbV^T \bbX$. So, if $\bbu^{(i,j)}$ is an eigenvector of $\bbS$ then we can not distinguish the values at nodes $i$ and $j$ and the problem remains non-identifiable. 

To exemplify this situation, consider the toy graph illustrated in Fig~\ref{fig:example_ambiguity}-(left). One can consider the adjacency matrix as the shift ($\bbS = \bbA$) and denote the corresponding eigenvectors as $\bbV = [\bbv_1,\cdots,\bbv_7]$. Fig~\ref{fig:example_ambiguity}-(center) shows that  $\bbv_4=\bbu^{(2,4)}$. Then it follows that for the matrix $\bbP$ in Fig~\ref{fig:example_ambiguity}-(right) and the vector $\bbp=[1,1,1,-1,1,1,1]^{T}$, one can construct another solution $\{\bbX_1,\tbh_1\}\neq\{\bbX_0,\tbh_0\}$ using \eqref{e:alternative_solution}. In other words, nodes $2$ and $4$ are indistinguishable. 

While it is challenging to obtain a formal characterization of problematic graphs, in practice we have encountered issues with dense networks as well as % with very sparse graphs such a trees.
with some very sparse graphs.
For (continuous-valued) weighted graphs such ambiguities effectively disappear. Before moving on to issues of exact recovery, a remark on identifiability of \eqref{e:blind_feasibility} for a simple but widely adopted (random) sparsity model is in order.

\begin{figure*}[t] % \label{f:figure_main}
	\begin{minipage}[b]{0.24\textwidth}
		\centering
		\includegraphics[width=1\linewidth]{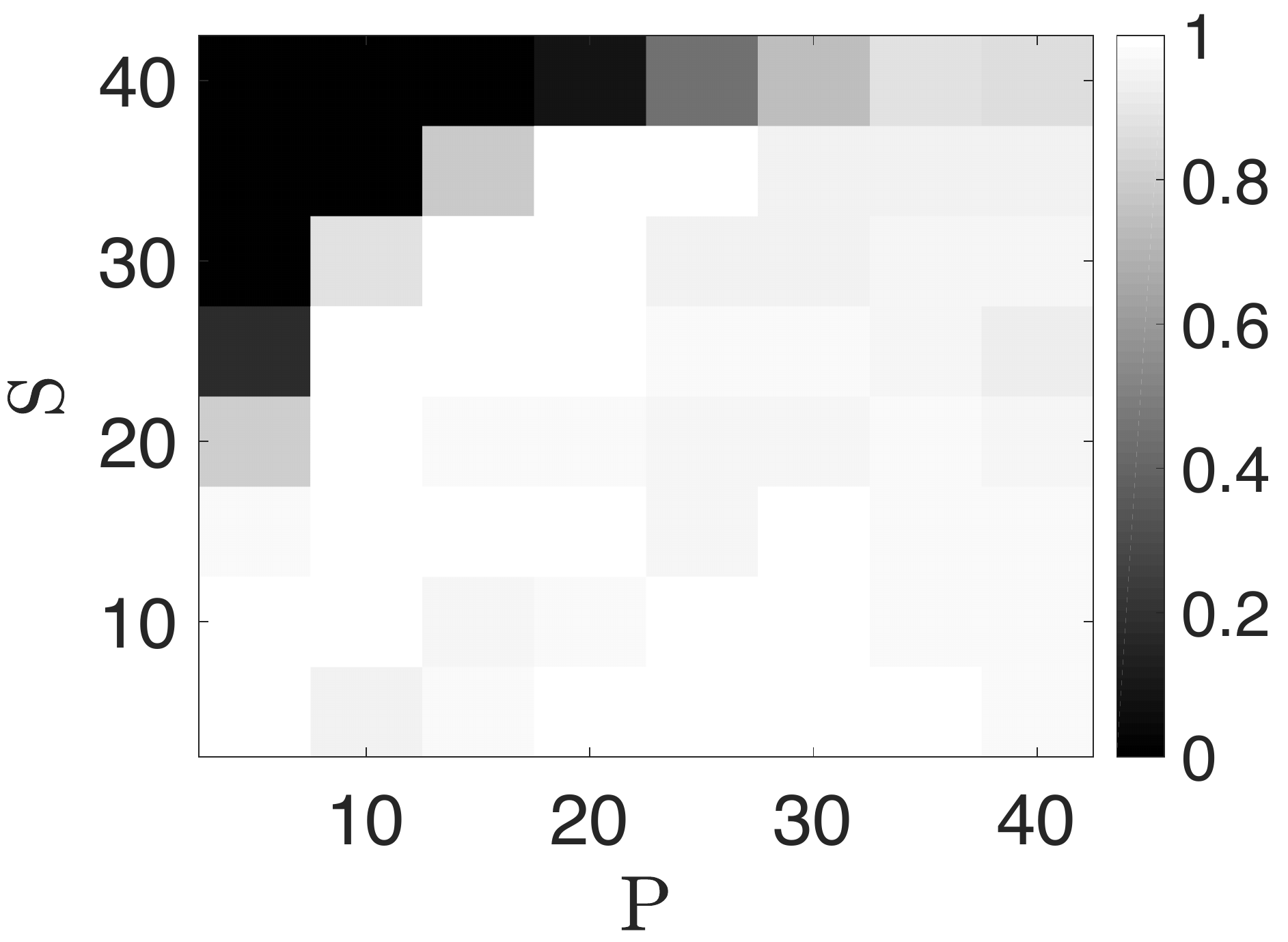}
		\centerline{(a)}\medskip
	\end{minipage}
	\hfill
	\begin{minipage}[b]{0.24\textwidth} 
		\centering
		\includegraphics[width=1\linewidth]{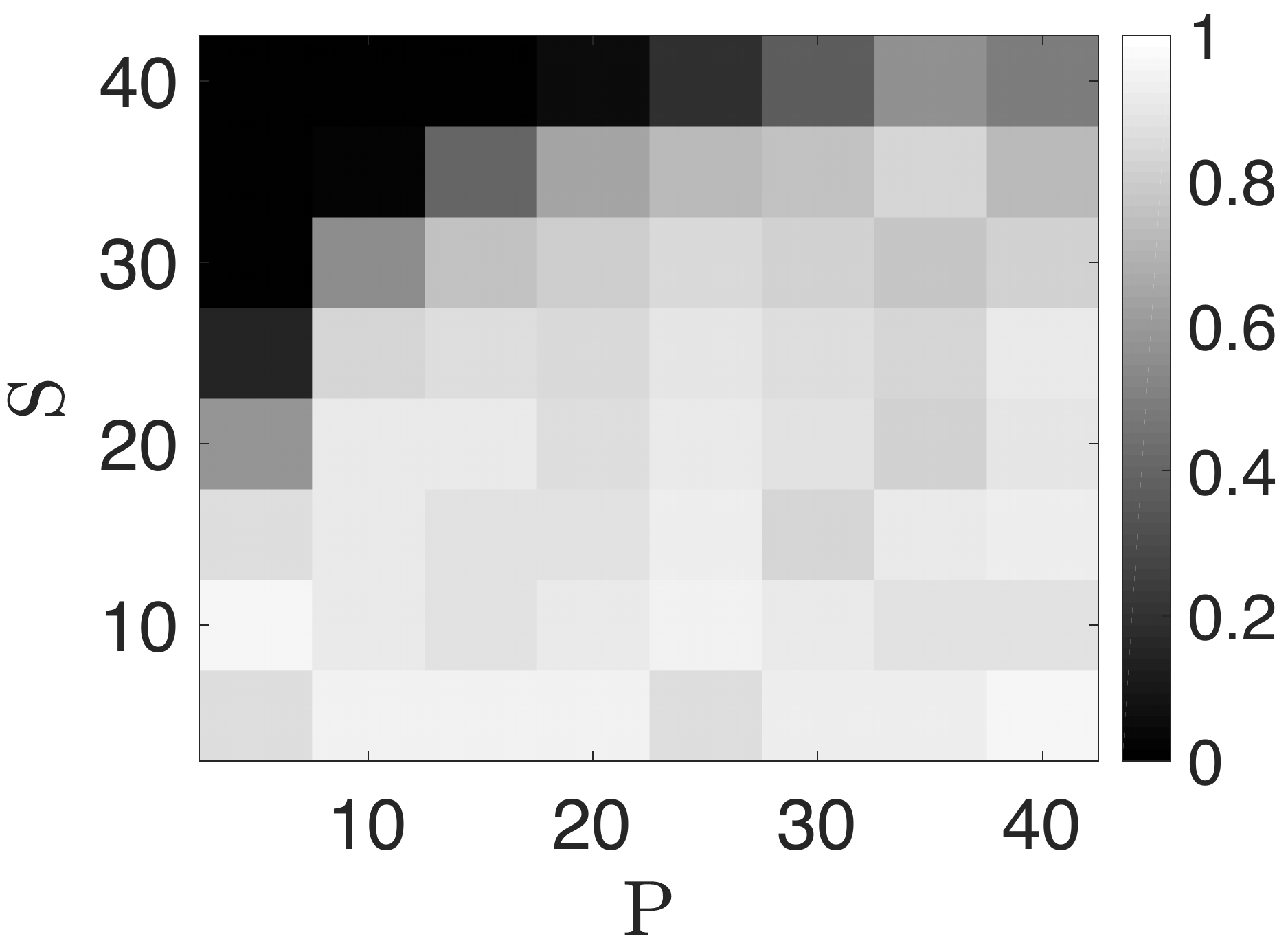}
		\centerline{(b)}\medskip
	\end{minipage}
	\hfill
	\begin{minipage}[b]{0.24\textwidth} 
		\centering
		\includegraphics[width=1\linewidth]{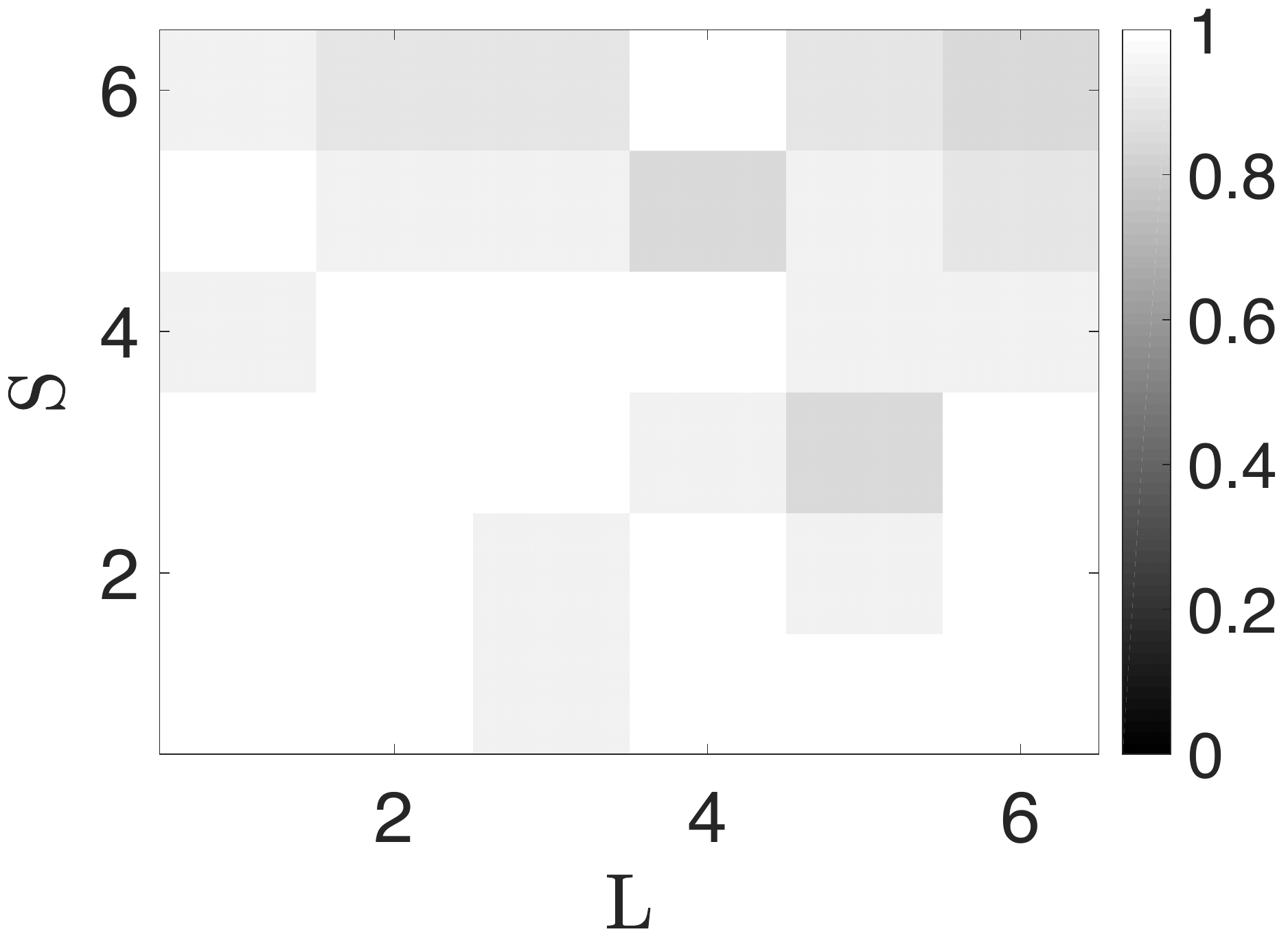}
		\centerline{(c)}\medskip
	\end{minipage}
	\hfill
	\begin{minipage}[b]{0.24\textwidth} 
		\centering
		\includegraphics[width=1\linewidth]{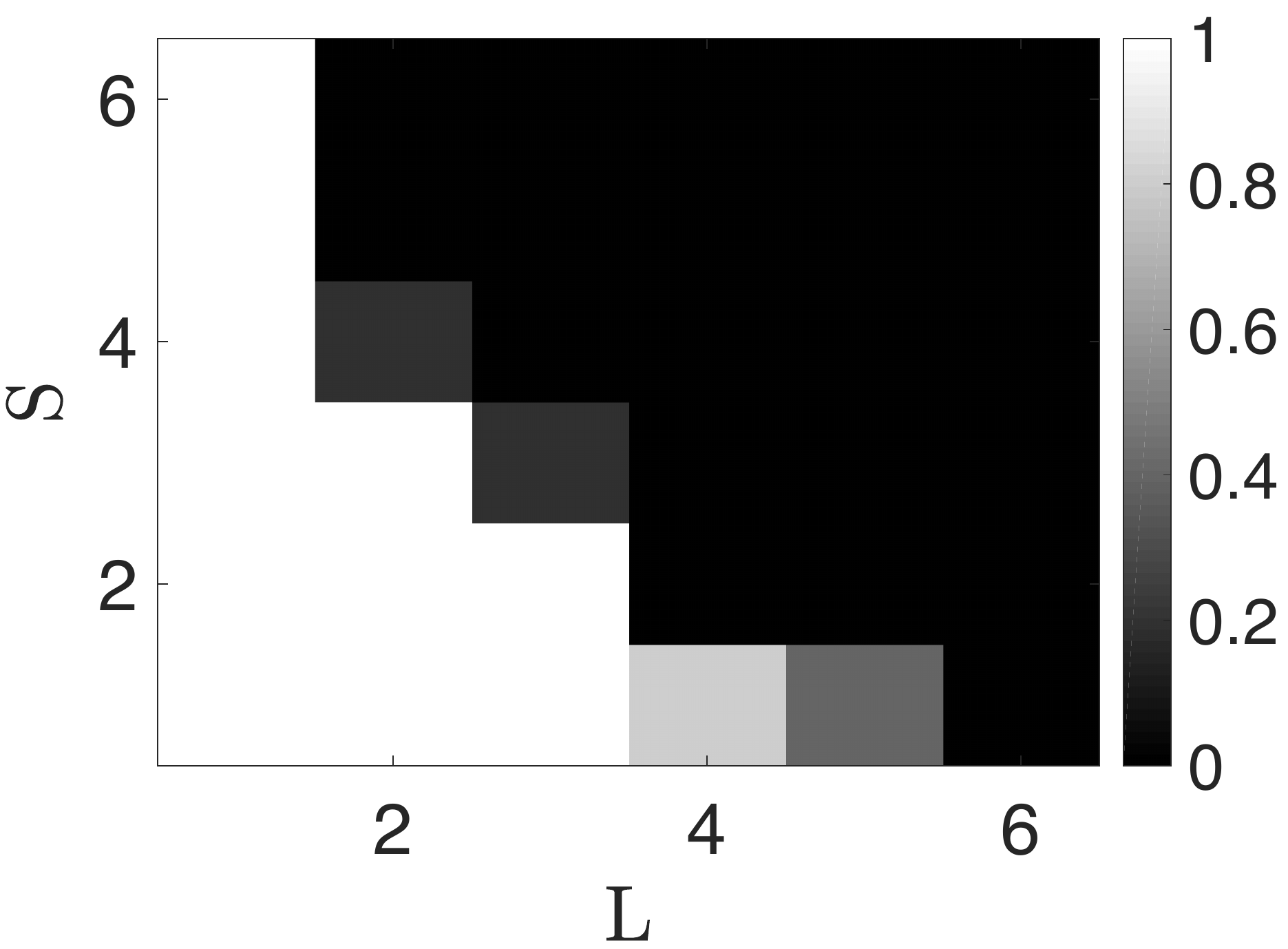}
		\centerline{(d)}\medskip
	\end{minipage}
	\vskip\baselineskip
	\vspace{-0.5cm}
	\begin{minipage}[b]{0.24\textwidth}
		\centering
		\includegraphics[width=1\linewidth]{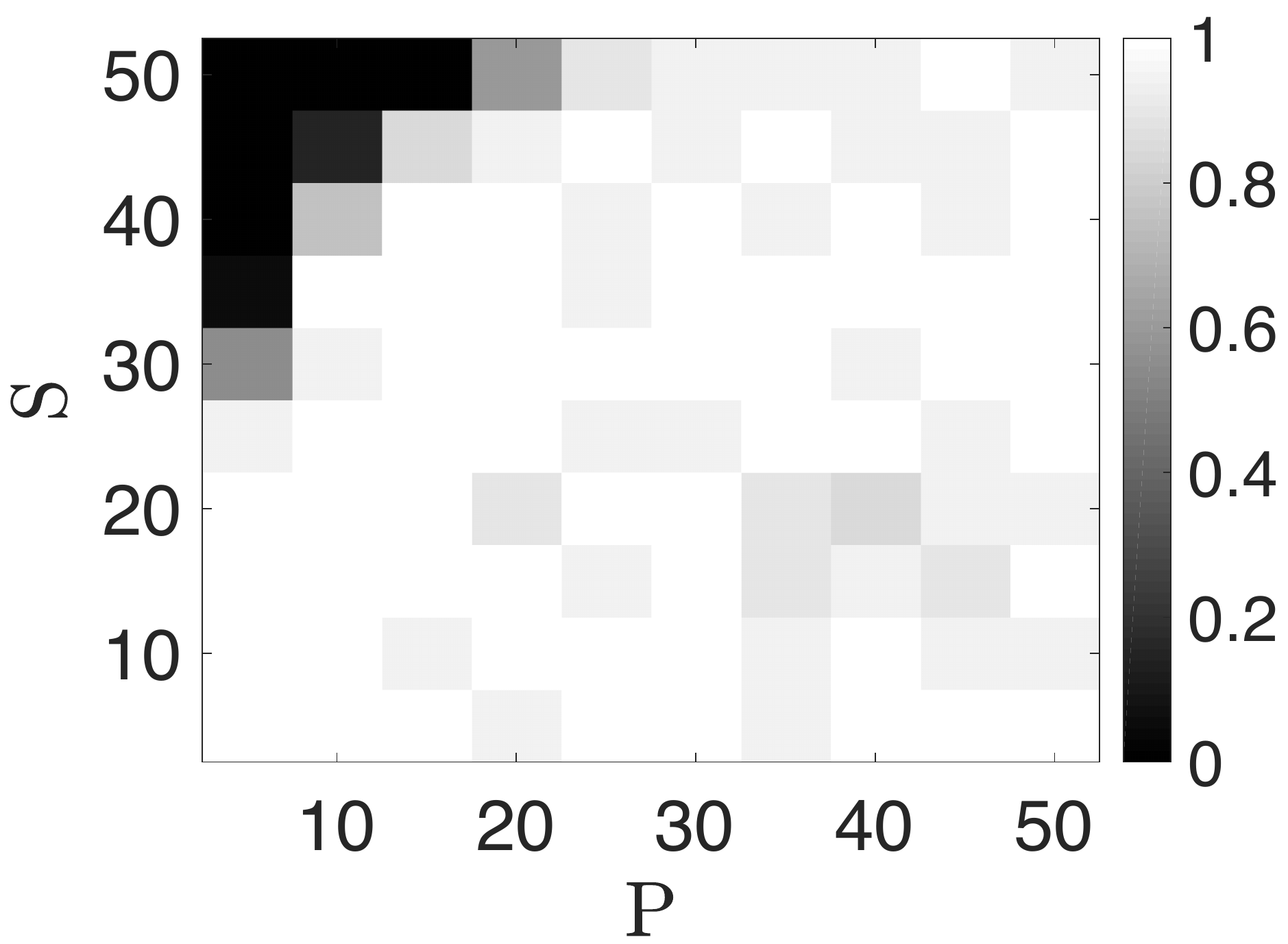}
		\centerline{(e)}\medskip
	\end{minipage}
	\hfill
	\begin{minipage}[b]{0.24\textwidth} 
		\centering
		\includegraphics[width=1\linewidth]{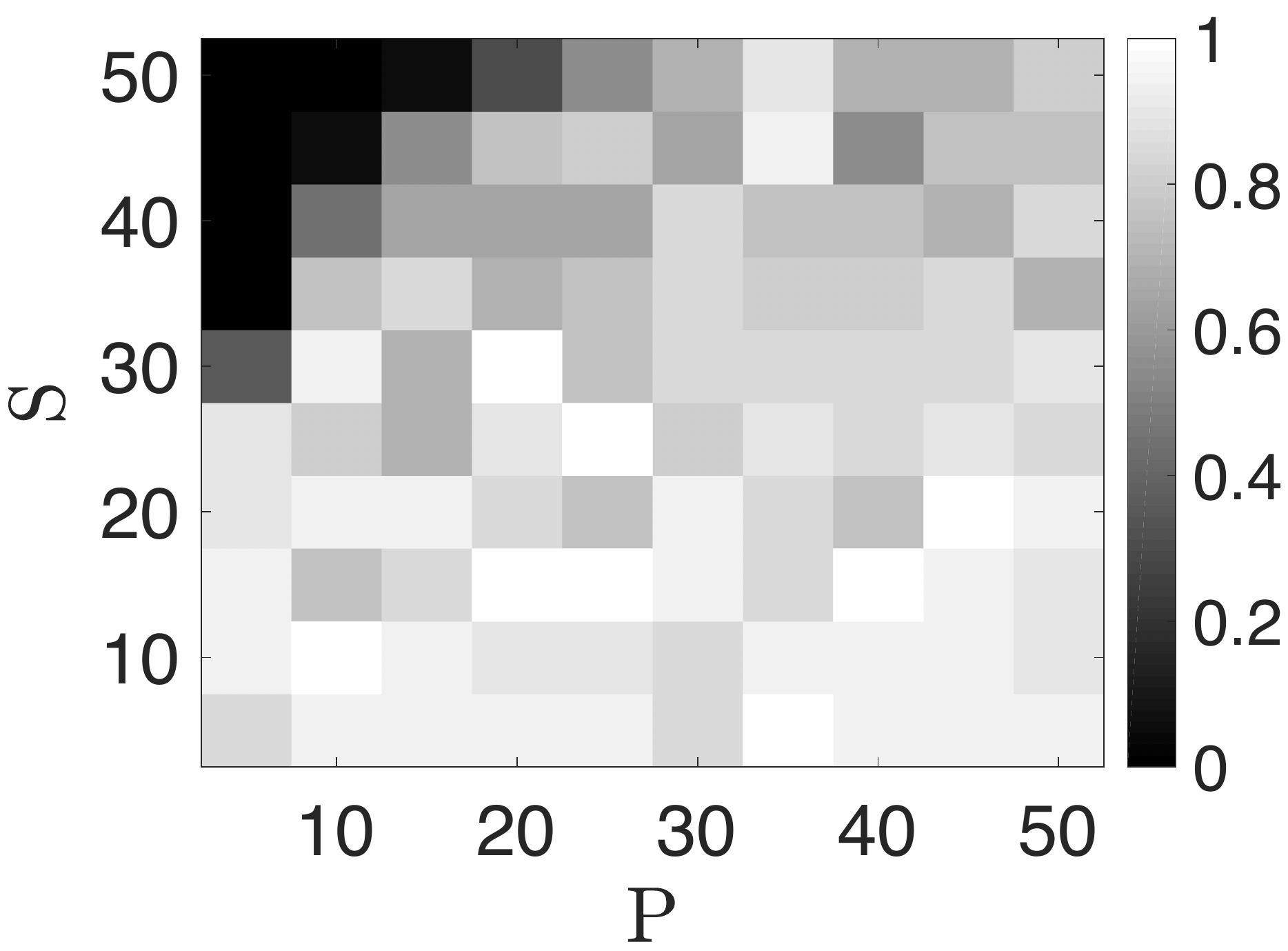}
		\centerline{(f)}\medskip
	\end{minipage}
	\hfill
	\begin{minipage}[b]{0.24\textwidth} 
		\centering
		\includegraphics[width=1\linewidth]{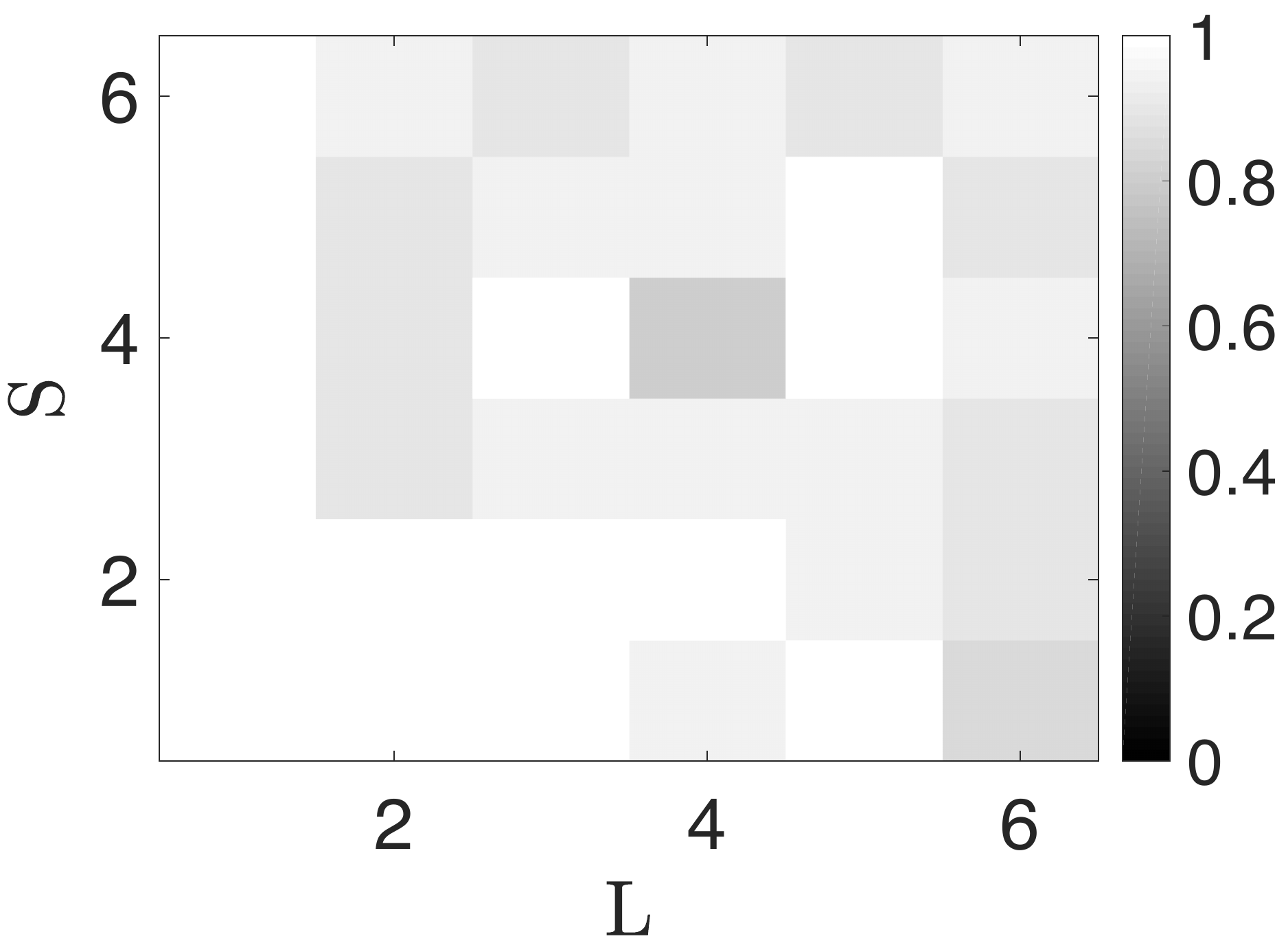}
		\centerline{(g)}\medskip
	\end{minipage}
	\hfill
	\begin{minipage}[b]{0.24\textwidth} 
		\centering
		\includegraphics[width=1\linewidth]{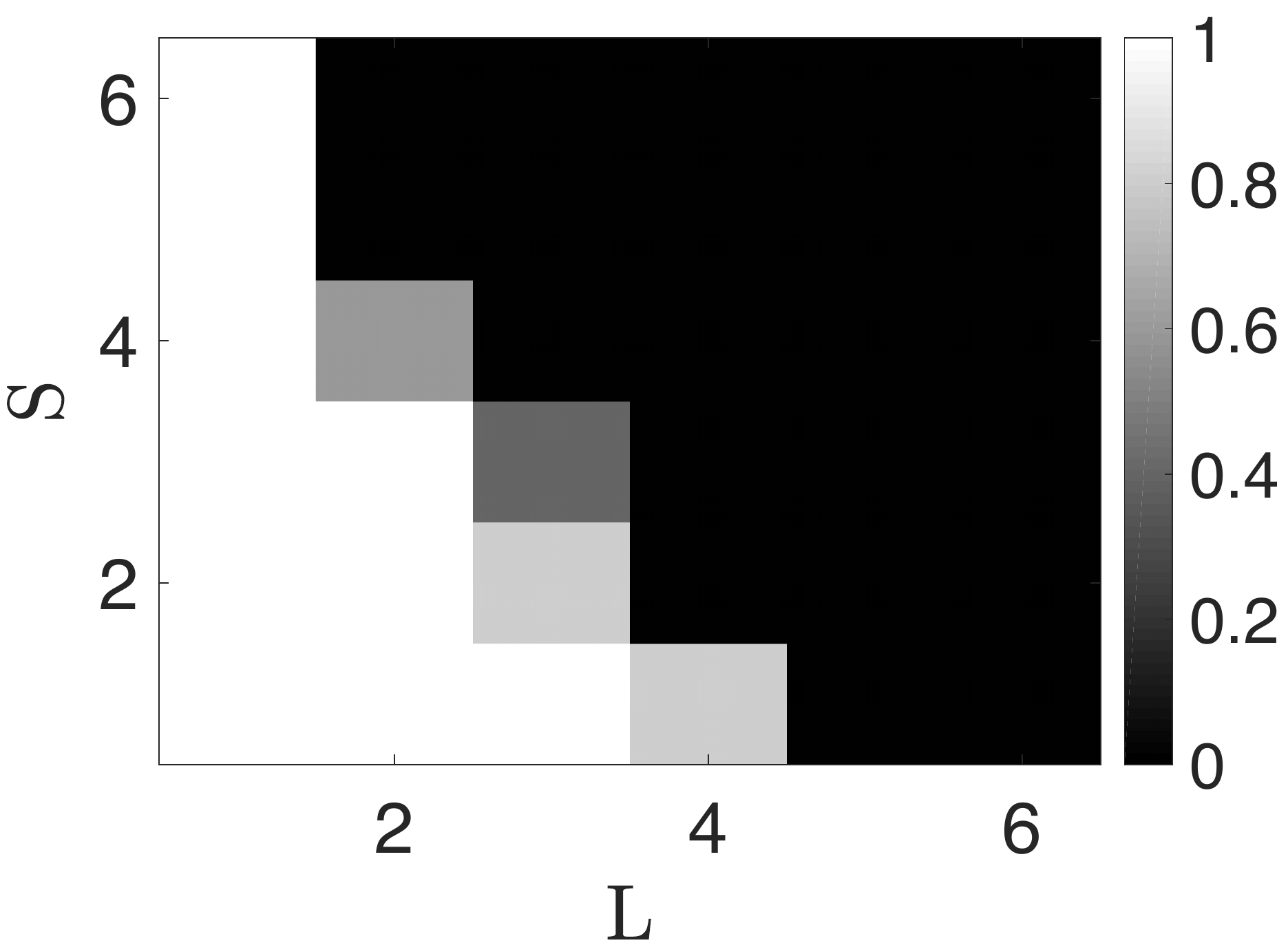}
		\centerline{(h)}\medskip
	\end{minipage}
	\vspace{-0.4cm}
	\caption{Rate of recovery of $\bbX$ as a function of $S$ (number non-zero entries in $\bbX$) and $P$ (number of observations) in $N=50$-node Erd\H{o}s-R\'{e}nyi graphs with $p=0.3$ (edge existence probability) for (a) $\alpha = 0.1$ and (b) $\alpha=0.3$, using    Algorithm~\ref{alg:reweighted_cvx}. Plots (e) and (f) are counterparts of (a) and (b), respectively, for the structural brain network in \cite{hagmann2008mapping}. Recovery rate in Erd\H{o}s-R\'{e}nyi graphs ($N=50$, $p=0.3$) as a function of $S$ and $L$ (filter order) using (c) Algorithm~\ref{alg:reweighted_cvx} and (d) the matrix-lifting approach of~\cite{segarra2017blind}. Plots (g) and (h) are counterparts of (c) and (d), respectively, for the aforementioned structural brain network.
	}
	\vspace{-0.3cm}
\end{figure*}

\begin{remark}[Identifiability for Bernoulli-Gaussian model] \normalfont \label{remark:bernouli_gaussian}
Because of its analytical tractability, the Bernoulli-Gaussian model is widely adopted to describe and generate random sparse matrices such as $\bbX \in \mathbb{R}^{N\times P}$ (we also use it for the simulations in Section~\ref{S:simulation}). Sparse matrices adhering to the model are $\bbX = \Omega \circ \bbR$, where $\Omega \in \reals^{N \times P}$ is an i.i.d. Bernoulli matrix with parameter $\theta$ (i.e., $\mathbb{P}[\Omega_{ij} = 1] = \theta$), and $\bbR \in \reals^{N \times P}$ is an independent random matrix with i.i.d. symmetric random variables drawn from a standard Gaussian distribution. Under the Bernoulli-Gaussian model, \cite[Proposition 40]{li2015unified}  asserts that problem \eqref{e:opt_blind_invertible} is identifiable (up to scaling and symmetric permutation ambiguities) with probability at least $1-\text{exp}(-c\theta P)$, for $\frac{1}{N}<\theta<\frac{1}{4}$ and  $P > cN\text{log}(N)$, where $c>0$ is a sufficiently large constant.
\end{remark}

\vspace*{-0.4cm}
\subsection{Exact recovery conditions} \label{S_s:unique}
 
 Suppose that \eqref{e:opt_blind_invertible} is identifiable and let $\{\bbX_0,\tilde{\bbg}_0 \}$ be the solution. The following proposition (that relies heavily on \cite[Theorem 1]{zhang2016one}) offers sufficient conditions under which the convex relaxation \eqref{e:opt_prob_convex} succeeds in exactly recovering $\{\bbX_0,\tilde{\bbg}_0 \}$.

%\vspace*{-0.3cm}
\begin{myproposition}\normalfont
Let $\ccalI:=\textrm{supp}(\textrm{vec}(\bbX_0))$ index the non-zero entries of  vectorized $\bbX_0$, and let $\ccalI^{c}$ be the complement of $\ccalI$. Moreover, define $\bbZ := \bbY^{T}\bbV \odot \bbV \in \reals^{NP \times N}$ and let $\bbZ_{\ccalS}$ be the submatrix of $\bbZ$ with rows indexed by $\ccalS\subset \{1,2,...,NP\}$. Then, the solution to \eqref{e:opt_prob_convex} is unique and equal to $\tilde{\bbg}_0$ if the two following conditions are satisfied:

\noindent \textbf{C1)} $\textrm{rank}(\bbZ_{\ccalI^c})=N-1$; and \\
\noindent \textbf{C2)} There exists a vector $\bbf\in\mathbb{R}^{NP}$ such that $\bbZ^T\bbf = \gamma \mathbf{1}_N$ for some $\gamma \neq 0$, such that $\bbf_{\ccalI}=\text{sign}(\bbZ_{\ccalI} \tilde{\bbg}_0)$ and $\Vert \bbf_{\ccalI^c}\rVert_{\infty}<1$. %, where  the infinity norm is denoted as $\lVert \bbx \lVert_{\infty}=\max_i |x_i|$.
\end{myproposition}

%\vspace*{-0.2cm}
\begin{myproof2}
As per \cite[Theorem 1]{zhang2016one}, $\tilde{\bbg}_0$ is the unique solution of \eqref{e:opt_prob_convex} if $\text{ker}(\bbZ_{\ccalI^c})\cap\text{ker}(\mathbf{1}_{N}) = \{\mathbf{0}\}$. But since $\tilde{\bbg}_0 \in \text{ker}(\bbZ_{\ccalI^c})$ and $\tilde{\bbg}_{0} \not \in \text{ker}(\mathbf{1}_{N})$ because of the constraint in \eqref{e:opt_prob_convex}, then C1) ensures said intersection is $\{\mathbf{0}\}$. Optimality condition C2) essentially requires $\mathbf{1}_N$ to belong to the set of subgradients of $\Vert \bbZ \tilde{\bbg}\rVert_{1}$ at $\tilde{\bbg}_0$; see \cite[Theorem 1]{zhang2016one} for further details.
\end{myproof2}

Naturally, a more insightful exact recovery and sample complexity result along the lines of the one in Remark \ref{remark:bernouli_gaussian}  would be most valuable [i.e., when are C1)-C2) satisfied for the Bernoulli-Gaussian model?], but left as future work.

\section{Numerical Results}\label{S:simulation}
We assess the performance of our proposed approach by testing the iteratively-reweighted $\ell_1$-norm minimization procedure in Algorithm \ref{alg:reweighted_cvx}. The per-iteration sparse recovery problems are solved using CVX~\cite{grant2008cvx}. 

\noindent\textbf{Simulation setup.} In all cases we consider undirected graphs with graph-shift operator chosen as the normalized adjacency matrix $\bbS = \bbD^{-\frac{1}{2}} \bbA \bbD^{-\frac{1}{2}}$, where $\bbD:=\textrm{diag}(\bbA \mathbf{1}_N)$ is a diagonal matrix of node degrees. The ground-truth sparse input matrix $\bbX_0$ is drawn from a Bernoulli-Gaussian model as in Remark~\ref{remark:bernouli_gaussian}, for varying parameters $N$, $P$, and sparsity level (i.e., number of nonzero entries) $S$. Filter coefficients $\bbh_0$ are generated according to 
$\bbh_0 = (\bbe_1 + \alpha \bbb)/\|\bbe_1 + \alpha \bbb\|_1$, where $\bbe_1=[1,0,\cdots,0]^T \in \reals^L$ is the first canonical basis vector and entries of $\bbb \in \reals^L$ are drawn independently from a standard Gaussian distribution. Such a model for $\bbh_0$ is inspired by \cite{wang2016blind}, and we later corroborate that the recovery performance improves as $\alpha$ decreases. Also note that $\bbh_0$ is normalized to unit $\ell_1$-norm to fix the scale of the problem. Finally, given $\bbX_0$ and $\bbH_0=\bbV\textrm{diag}(\bbPsi_L\bbh_0)\bbV^T$, the $N\times P$ matrix of observations is generated as $\bbY = \bbH_0 \bbX_0$. 

The relative recovery error $e_X = \|\hat{\bbX}-\bbX_{0}\|/\|\bbX_{0}\|$ is adopted as figure of merit to evaluate algorithmic performance. We estimate the rate of successful recovery for synthetic and real-world graphs under different parameters by defining a successful recovery as one with $e_X < 0.01$.

\noindent\textbf{Random graphs.} Consider Erd\H{o}s-R\'{e}nyi random graphs with $N = 50$ nodes, where edges are formed
independently with probability $p = 0.3$. The rate of successful recovery is estimated for realizations of random graphs which are connected and do not give rise to permutation ambiguities (cf. Section~\ref{S_s:per}). Figures 2(a) and 2(b) depict the recovery rates as a function of $P$ and $S$ for $\alpha = 0.1$ and $0.3$, respectively, averaged over $100$ realizations for (invertible) graph filters of order $L=5$. As expected, in both cases recovery is more challenging for larger $S$ and smaller $P$; see the dark-gray area of low-success probability around the top-left corner. Moreover, decreasing $\alpha$ makes successful recovery more likely. For instance, for $\alpha=0.1$ [Fig. 2(a)] we can successfully recover dense input signals with e.g., $S \approx N/2 = 25$ and only $P = 10$ observations. For (effectively) lower-order filters resulting in more localized diffusion dynamics, one obtains favorable recovery performance.

We also compare the proposed approach against its less-scalable, matrix lifting-based precursor in \cite{segarra2017blind}. %Remarkably, Algorithm~\ref{alg:reweighted_cvx} yields an order-of-magnitude speedup over state-of-the-art  \cite{segarra2017blind}. 
Figures 2(c) and 2(d) respectively show the recovery rates for both methods  as a function of sparsity $S$ and filter order $L$, for $N=50$, $p=0.3$, $P = 10$, and $\alpha = 0.5$ averaged over $20$ realizations. Apparently, Algorithm~\ref{alg:reweighted_cvx} [Fig. 2(c)] can be successful over a larger range of values of $L$. Moreover, it uniformly outperforms the  algorithm in~\cite[Problem (9)]{segarra2017blind}; see Fig. 2(d).

\noindent\textbf{Brain graph.} We also consider a structural brain graph with $N=66$ nodes or neural regions of interest (ROIs), and edge weights given by the density of anatomical connections between regions~\cite{hagmann2008mapping}. The level of activity of each
ROI can be represented by a graph signal $\bbx$, thus successive applications of $\bbS$ model a linear evolution of the brain activity pattern.
Supposing we observe a linear combination (filter) of the evolving states of an originally sparse brain signal, then blind identification
amounts to jointly estimating which regions were originally active, the activity in these regions and the coefficients of the linear combination.

We repeat the recovery-rate analysis performed for Erd\H{o}s-R\'{e}nyi graphs averaged over $20$ realizations, and report the results in Figs. 2(e)--(h). Figures 2(e) and 2(f) showcase that our algorithm successfully identifies the initial excitation regions as well as the diffusion coefficients over a broad region in parameter space. By comparing Figs. 2(g) and 2(h), it is apparent that also in this setting the proposed approach outperforms the state-of-the-art method in~\cite{segarra2017blind}, corroborating the effectiveness of  Algorithm~\ref{alg:reweighted_cvx}.

\vspace*{-0.2cm}
\section{Conclusion}\label{S:conclusion}

We studied the problem of blind graph filter identification, which extends blind deconvolution of time (or spatial) domain signals to
graphs. By introducing a mild assumption on invertibility of the graph filter, we obtained a computationally simpler convex relaxation for (diffused) source localization in the multi-signal case. Ongoing work includes deriving suitable graph-dependent conditions under which exact (and stable) recovery can be guaranteed, even when only a fraction of nodes is observed. This is a challenging problem, since the
favorable (circulant) structure of time-domain filters is no longer present in the network-centric setting dealt with here.

%\urlstyle{same}
\bibliographystyle{IEEEtranS}
\vspace*{-0.2cm}
\bibliography{citations}
\end{document}